\documentclass[preprint,showpacs,preprintnumbers,nofootinbib,
superscriptaddress]{revtex4}
\usepackage{graphicx}
\usepackage{epsfig}
\usepackage{dcolumn}
\usepackage{bm}

\def\gsim{\lower0.5ex\hbox{$\:\buildrel >\over\sim\:$}}
\def\lsim{\lower0.5ex\hbox{$\:\buildrel <\over\sim\:$}}

\newcommand{\bea}{\begin{eqnarray}}
\newcommand{\eea}{\end{eqnarray}}

\begin{document}

\preprint{HIP-2005-03/TH}
\preprint{TIFR/TH/05-05}

\title{Infrared fixed point of the top Yukawa coupling 
in split supersymmetry}

\author{Katri Huitu}
\email{Katri.Huitu@helsinki.fi}
\affiliation{High Energy Physics Division, Department of Physical
Sciences, P.O. Box 64, FIN-00014 University of Helsinki, Finland}
\affiliation{Helsinki Institute of Physics, 
P.O. Box 64, FIN-00014 University of Helsinki, Finland}
\author{Jari Laamanen}
\email{jalaaman@pcu.helsinki.fi}
\affiliation{High Energy Physics Division, Department of Physical
Sciences, P.O. Box 64, FIN-00014 University of Helsinki, Finland}
\affiliation{Helsinki Institute of Physics,        
P.O. Box 64, FIN-00014 University of Helsinki, Finland}
\author{Probir Roy}%
\email{probir@theory.tifr.res.in}
\affiliation{Department of Theoretical Physics, Tata Institute of Fundamental 
Research, Homi Bhabha Road, Mumbai - 400 005, India}
\author{Sourov Roy}
\email{roy@pcu.helsinki.fi}
\affiliation{Helsinki Institute of Physics,        
P.O. Box 64, FIN-00014 University of Helsinki, Finland}

\received{\today}

\pacs{12.60.Jv, 14.65.Ha, 14.80.Ly} 

\begin{abstract} \vspace*{10pt}
The severe constraints imposed on the parameter space of the minimal
split supersymmetry model by the infrared fixed point solution of the
top Yukawa coupling $Y_t$ are studied in detail in terms of the value
of the top quark mass measured at the Tevatron together with the lower
bound on the lightest Higgs mass established by LEP. The dependence 
of the higgsino mass parameter $\mu$, the gaugino coupling strengths 
${\tilde g}_{u,d}$, ${\tilde g}^\prime_{u,d}$ and of the Higgs quartic
self coupling $\lambda$ on the value of $Y_t$ in the vicinity of the 
Landau pole is discussed. A few interesting features emerge, though the model
is found to be disfavored within the infrared fixed point scenario
because of the need to have several unnatural cancellations at work on 
account of the requirement of a low upper bound on $\tan\beta$.
\end{abstract}

\maketitle

\section{Introduction}
The naturalness criterion has been one of the guiding principles in the
formulation of the (Minimal) Supersymmetric Standard Model (M)SSM. Once
this is accepted, a successful implementation of high scale gauge
coupling unification obtains and the Lightest Supersymmetric Particle
(LSP) emerges as a viable dark matter candidate. But, in view of the
failure of the above criterion in dealing with the cosmological constant
and in the light of the recently advanced landscape paradigm, an
important question arises. Can one abandon the principle of
naturalness, admit fine tuning and yet maintain the nice phenomenological 
aspects of the SSM at the same time? It has been emphasized \cite{arkani-dimo,
giudice-romanino} that the successful unification of gauge couplings of
the SSM can be retained even when all the scalars of the theory, except
one finetuned light Higgs boson (akin to that in the Standard Model)
lie far above the electroweak scale. Thus, despite the loss of the
original motivation to cure the hierarchy problem, one can still have a
supersymmetric theory with gauge coupling unification, which is free of
many of the undesirable features of the SSM such as the flavor problem,
fast proton decay via dimension five operators, generically large CP
violation, a tightly constrained mass of the lightest Higgs etc. The
gauginos and higgsinos of this theory are chosen to lie near the TeV
scale to ensure gauge coupling unification at ${\rm M_{GUT}} \sim 10^{16}$
GeV  as well as a stable LSP in the desirable mass range. This is the
scenario of split supersymmetry, as named in Ref.
\cite{giudice-romanino}.  

Various theoretical and phenomenological aspects, characteristic of the
above scenario, have been discussed in several recent works 
\cite{dark-matter, one-loop, aspects, spectrum, collider, zhu,2-loop, axion, 
variation, r-parity, neutrino, neutralino-dark, diaz-perez, hewett,
cosmic, keung, sarkar, allahverdi, bajc, graham, senatore, vacuum,
schuster, gao, martin, chen, babu, drees, shinta, cheung}. One can 
identify a minimal split supersymmetry model described by six 
specific parameters : (1) a common mass $\widetilde m$ for the heavy 
scalars, (2) $\tan\beta$, where the angle $\beta$ defines the combination of 
neutral ${\rm SU(2)_L}$-doublet Higgs fields which remains light, (3) 
the higgsino mass parameter $\mu(M_{GUT})$ at the GUT scale, (4) the 
gluino mass $m_{\tilde g}$, (5) the grand unification scale 
${\rm M_{GUT}}$, and (6) the unified value of the gauge coupling 
strength $\alpha_G$ at ${\rm M_{GUT}}$. However, 
the last two are more or less fixed by the requirement of consistency with
measurements of the three gauge coupling strengths at laboratory
energies. It is thus convenient to discuss different phenomenological
constraints in the space of the first four parameters. It has been
already realized \cite{giudice-romanino} that certain special
constraints would ensue (on the parameter space of the minimal 
split SUSY model, in particular) on account of the Landau pole 
\cite{hill,ibanez} in the top quark Yukawa coupling $Y_t$ and the LEP 
lower bound on the mass of the Standard Model Higgs. However, a careful 
quantitative study of those, including the interrelation between the 
last mentioned two aspects, has been lacking and that is the aim of the 
present work.

We broadly embrace the philosophy of Refs.\cite{giudice-romanino} and 
\cite{aspects} in this paper. Our gluino and electroweak gaugino as well
as higgsinos are envisioned to lie in the range of hundreds of GeV
whereas $\widetilde m$ is taken to be much above 10 TeV and most likely 
around $10^9$ GeV. Indeed we vary $\widetilde m$ all the way upto 
$10^{13}$ GeV beyond which scale one might encounter anomalously heavy 
isotopes \cite{giudice-romanino}. We follow the RGE equations set up in
Ref.\cite{giudice-romanino} and numerically study the parameters of the
minimal split SUSY model as $\widetilde m$ is varied with $Y_t$ kept 
at its fixed point value or in its vicinity. Since the higgsino mass 
parameter $\mu(M_Z)$ and the gaugino couplings are sensitive to 
values of $Y_t$ in this region, we study them as functions of the top 
mass $m_t$ with $\widetilde m$ fixed. In Section II we first review the 
physics of the infrared fixed point of $Y_t$ in MSSM and then extend the 
discussion to split supersymmetry. In Section III we consider the
implications of this scenario for the Higss mass $M_h$, the higgsino mass
parameter $\mu$ as well as the gaugino coupling strengths. Section IV 
contains our conclusion and the RGEs are relegated to the Appendix.

\section{Infrared fixed point of $Y_t$}
Let us first review the fixed point behaviour \cite{hill} of the top 
Yukawa coupling in MSSM. In the low to moderate $\tan\beta$ region, the
effects of the bottom and tau Yukawa coupling strengths can be ignored.
With this approximation and, given gauge coupling unification at ${\rm
M_{GUT}}$, one obtains a simple analytic relation \cite{ibanez,savoy,
carena,nath,kobayashi,biswajoy} at the one-loop level : 
\bea
Y_t (t) = \frac {Y_t(0) E(t)} {1 + 6 F(t) Y_t (0)}.
\label{eqn1}
\eea
\noindent In Eqn. (\ref{eqn1}), $t = 2~{\rm ln}({M_{GUT}}/Q)$, $Y_t =
{\lambda^2_t} / {{(4 \pi)}^2}$, $\lambda_t$ is the top Yukawa 
coupling strength in the Lagrangian, $Q$ is the running scale variable,
$E$ and $F$ are functions of the gauge couplings:
\bea
E(t) = {(1 + \beta_3 t)}^{16/{3b_3}} {(1 + \beta_2 t)}^{3/{b_2}} {(1 +
\beta_1 t)}^{13/{9 b_1}}, ~~~~F(t) = \int_0^t E(t^\prime)dt^\prime.
\label{eqn2}
\eea  
\noindent The parameters $\beta_i\,(i=1,2,3)$ in Eqn.(\ref{eqn2}) equal 
${b_i \alpha_G}/{(4\pi})$, where ($b_1, b_2, b_3$) = (33/5, 1, -3) are
the coefficients of the one-loop gauge $\beta$-function and 
$\alpha_G = \alpha_i(0)$ with the normalization $\alpha_1 = {\frac 5
3}\alpha_Y$ for the hypercharge coupling. Eqn.(\ref{eqn1}) implies that
a large value ($\sim 3.5$) at the GUT scale of the top Yukawa coupling
$\lambda_t$ in the Lagrangian corresponds to an infrared quasi-fixed
point value of $Y^f_t$ :
\bea
Y^f_t(t) = E(t)/{6F(t)}.
\label{eqn3}
\eea

The situation is somewhat different in split supersymmetry. 
Here all sfermions and the charged as well as the heavier CP even plus
the CP odd Higgs bosons are very heavy and, as a first approximation,
are taken to be degenerate\footnote{Non-universal scalar masses in the
split supersymmetry scenario have been considered \cite{non-univ}.}. 
Coupling strengths in the split theory at the scale $\widetilde m$ are 
obtained by matching its Lagrangian with that of the full MSSM
valid at higher scales. In particular, the couplings of the light Higgs 
$h$ in the split effective theory follow from matching conditions with 
the interaction terms of the Higgs doublet fields $H_u$ and $H_d$ in the 
full MSSM. Suppose we denote the top Yukawa coupling strength in the 
Lagrangian of the effective theory as $h_t$. If $\lambda_t$ represents 
the coupling strength of the Yukawa interaction of the top with $H_u$ 
in the full MSSM above $\widetilde m$, then we have
\cite{giudice-romanino}
\bea
h_t (\widetilde m) = \lambda^*_t(\widetilde m)\sin\beta,
\label{eqn4}
\eea

The evolution of $\lambda_t$ at scales greater than $\widetilde m$ is given
at the one-loop level by Eqn. (\ref{eqn1}). However, below the scale
$\widetilde m$, $h_t$ evolves according to Eqn.(\ref{eq-a.13}) given in the
Appendix with the matching condition of Eqn.(\ref{eqn4}). With this
evolution also, an infrared fixed point is observed for $h_t = h^f_t$.
Though an analytic expression for $h^f_t$ becomes complicated, this
striking behaviour can be seen numerically. The corresponding top quark
pole mass is then given by \cite{polemass}
\bea
M^{pole}_t = h_t^f (M_Z) v\left[1 + {{4\alpha_3(M_Z)} \over {3 \pi}} 
-2Y^{\prime f}_t(M_Z)\right],
\label{eqn5} 
\eea
$v$ being $\approx$ 246 GeV and $Y^{\prime f}_t = {(h^f_t)}^2/ 
{(4\pi)}^2$. In our numerical calculations we have also taken into
consideration the effects of bottom and tau Yukawa couplings.  

\begin{figure}
\includegraphics{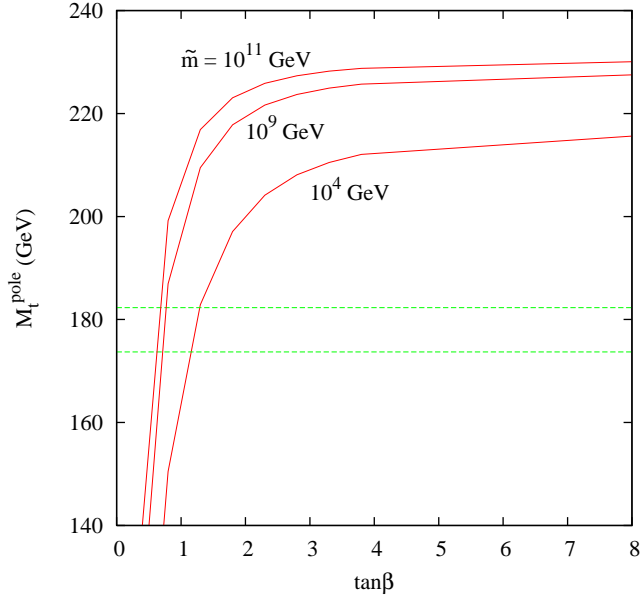}
\caption{\label{fig1}
Top pole mass at the infrared fixed point value as a function of 
$\tan\beta$ for three different values of $\widetilde m$ with the 
1-$\sigma$ band of the Tevatron measurement also shown.}
\end{figure}

\begin{figure}
\includegraphics{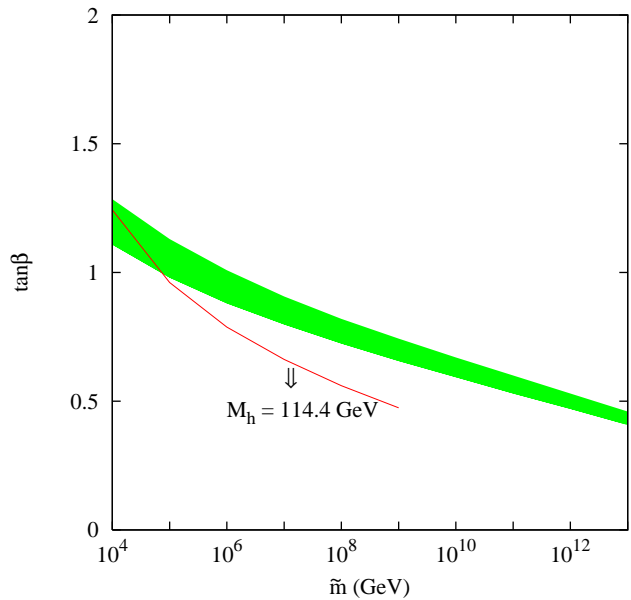}
\caption{\label{fig2}
Allowed region (coloured) in the ${\widetilde m}$ -- $\tan\beta$ plane 
in the infrared fixed point scenario from the experimental limits on 
the top mass. The area below the solid line is disallowed by the 
LEP--2 lower limit \cite{lephiggs} of $M_h >$ 114.4 GeV. $M_{1/2}$ and
$\mu$ at the GUT scale have been chosen at 300 GeV and --600 GeV 
respectively.}
\end{figure}

In split supersymmetry $\tan\beta$ enters as an input parameter into the
top mass via Eqn.(\ref{eqn4}). The experimental upper (lower) limit on
the top mass then translates to an upper (lower) limit on $\tan\beta$.
This feature is demonstrated in Fig.~\ref{fig1} for three values of
$\widetilde m$, namely $10^4$ GeV, $10^9$ GeV and $10^{11}$ GeV. We have
calculated the results numerically upto $\tan\beta$ = 40 but plotted
them only in the small $\tan\beta$ region which is the most interesting part
to look for in this context. The values of $\mu$ and $M_{1/2}$ at the GUT 
scale have been taken here to --600 GeV and 300 GeV, respectively. 
The 1$\sigma$ error in $M_t^{pole}$, as currently quoted in the PDG 
listing \cite{PDG}
\bea
M_t^{pole} = 178.0 \pm 4.3~{\rm GeV},
\label{eqn6}
\eea 
in combination with its infrared fixed point value, 
puts bounds on $\tan\beta$ defined at the scale $\widetilde m$.  
An interesting new feature, different from what happens in the MSSM,
is that $\tan\beta$ can now be lower than unity for large values of
$\widetilde m$.  However, the most important point is that the fixed point
value of the top mass is now consistent with only a thin sliver of an
allowed region in the $\tan\beta-\widetilde m$ plane, as shown in
Fig.\ref{fig2}. On the other hand, if we do not stick to the fixed
point scenario, this severe restriction weakens considerably though a
lower bound on $\tan\beta$ continues to exist and is correlated to the
lower limit on $M_t^{pole}$. 

The value of $\tan\beta$ in models of split
supersymmetry depends upon \cite{arkani-dimo,drees} what one assumes for
the strength of the $B$-parameter, but it is generally difficult to keep
$\tan\beta$ small. If $|B|$ is of the order of the EW symmetry
breaking scale $m_{EW}$ then ${\tan\beta} \sim {{\widetilde m}^2 /
{m^2_{EW}}} >$ 100 for ${{\widetilde m} / {m_{EW}}} >$ 10, violating the
upper bound $\lsim$ 100 on $\tan\beta$ coming from the need to keep the
bottom Yukawa coupling strength perturbative, i.e. $\lsim {\cal O}(1)$.
On the other hand, in usual gravity-, gauge- or anomaly-mediated
supersymmetry breaking, it is possible to have $|B|$ of the order of
$\widetilde m$. In this case, one has ${\tan\beta} \sim {{\widetilde m} /
{m_{EW}}}$ which allows somewhat larger splitting in the spectrum while
keeping the value of $\tan\beta$ within the above-mentioned upper limit. 
However, it is still not sufficient to ensure that $\tan\beta$ remains
within the allowed region of Fig. \ref{fig2}. We have just seen that in the 
infrared fixed point scenario in split supersymmetry the upper bound on 
$\tan\beta$ (as a function of $\widetilde m$) is very strong ($\tan\beta 
\lsim$1 for large values of $\widetilde m$). Thus, combining this observation 
with the above argument one can perhaps conclude that the infrared fixed 
point scenario is strongly disfavored in split supersymmetry in the 
context of gravity-, gauge- or anomaly-mediated supersymmetry breaking 
(with $|B| \sim {\widetilde m}$ or in the case when $|B| \sim m_{EW}$). 
In other words, if $\tan\beta$ is experimentally measured to be $\lsim$ 1 
with a sparticle spectrum that contains physical charginos and neutralinos 
but with the scalars (except for one light Higgs) being out of the LHC 
energy reach, the infrared fixed point scenario can probably be retained 
but either at the cost of several unnatural cancellations having to work 
together \cite{drees} or having a direct mediation mechanism with D-term 
supersymmetry breaking ($|B| \gg {\widetilde m}$ and\footnote{Recall that the 
Higgs mass mixing term $B\mu$ needs to be of the same order as 
${\widetilde m}^2$.} $|\mu| \ll {\widetilde m})$~ which introduces additional 
heavy matter fields or a new scale in the theory \cite{aspects, babu}.

\section{Implications of fixed point for other masses and couplings}
Let us now study how the light Higgs mass $M_h$ changes with $\tan\beta$
when the top mass is at its fixed point value. As in the Standard Model,
$M_h$ in split supersymmetry can be written as 
\bea
M_h = \sqrt{\lambda}v,
\label{eqnhmass}
\eea
\noindent where $\lambda$ is the strength of the quartic self-coupling
of $h$, and $v$ is as in Eqn.(\ref{eqn5}).  The matching condition for
the coupling $\lambda$ at the scale $\widetilde m$ is 
\bea
\lambda({\widetilde m}) = {\frac {[g^2 ({\widetilde m}) 
+ g^{\prime 2}({\widetilde m})]} {4}} {\rm cos}^2 2\beta,
\label{eqnlambda}
\eea 
\noindent where $g$ and $g^\prime$ are the respective $SU(2)_L$ and 
$U(1)_Y$ coupling strengths with $\alpha_1 = 5 g^{\prime 2}/
({12 \pi})$. The evolution of $\lambda$ is governed by
Eqn.(\ref{eq-a.24}) of the Appendix. The mass $M_h$ also constrains 
$\tan\beta$ as a function of $\widetilde m$, cf. Fig. \ref{fig2}. 

\begin{figure}
\includegraphics{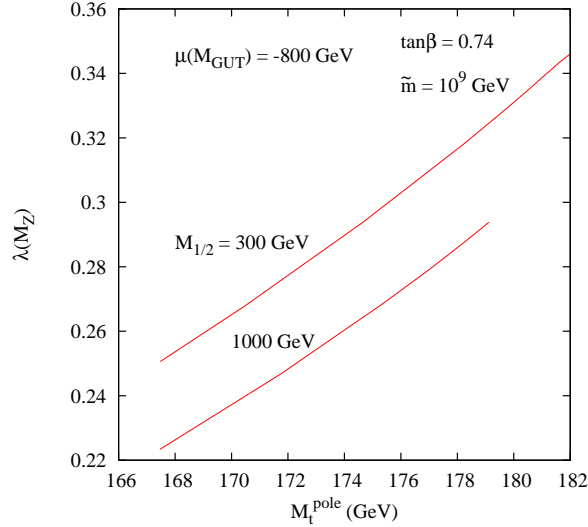}
\caption{\label{fig3}
Variation of the quartic coupling $\lambda(M_Z)$ near the top-quark fixed 
point for two values of $M_{1/2}$. Here, $\widetilde m = 10^9$ GeV and 
$\tan\beta$= 0.74. The value of $\mu(M_{GUT})$ is taken to be -800 GeV.}
\end{figure}

It is also interesting to note how the quartic coupling $\lambda(M_Z)$
changes with the top mass near the fixed point value. In Fig. \ref{fig3}
we have shown this variation for a fixed $\tan\beta$ and $\widetilde m$
and for two values of the common gaugino mass $M_{1/2}$. In both the
cases the fixed point value of the top mass is within the 1$\sigma$
limit given in Eqn. \ref{eqn6}. We can see from this figure that 
$\lambda(M_Z)$ shows some variation with the top mass near the fixed 
point. Accurate knowledge of chargino and neutralino masses (which will
determine $M_{1/2}$) and of the top mass will enable one to obtain a
precise value of $\lambda(M_Z)$ and then one can calculate the value of
$\lambda(\widetilde m)$ using the split susy RGE and verify the
prediction given in Eqn.\ref{eqnlambda}. This figure is plotted for a
fixed value of $\mu(M_{GUT})$ = -800 GeV but we have checked that the
variation of $\lambda(M_Z)$ with $M_t$ does not have any significant
dependence on $\mu(M_{GUT})$ by varying the latter between -800 GeV and
+800 GeV\footnote{In split supersymmetry, the neutralino and chargino 
masses (and hence $|\mu(M_Z)|$) cannot be much higher than $\cal O$(TeV). 
The latter requirement, together with the extremely small region of 
$\tan\beta$, i.e. 0.5 $< \tan\beta <$ 1.3 (cf. Fig.\ref{fig2}), allowed 
in the infrared fixed point scenario, means that here a $|\mu(M_{\rm GUT})|$,
much larger than $\cal O$(TeV), is disallowed since it will not be able to
run down to an acceptable value of $|\mu(M_Z)|$.}.

\begin{figure}
\includegraphics{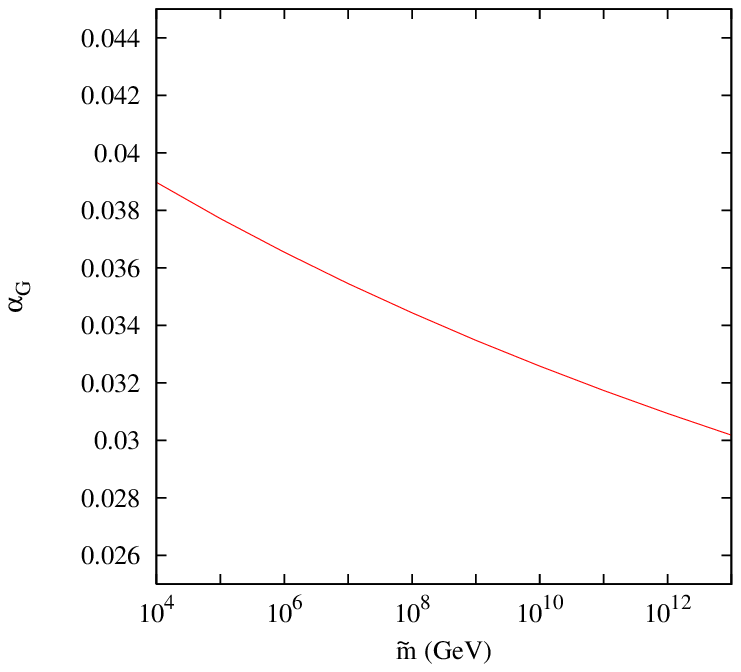}
\includegraphics{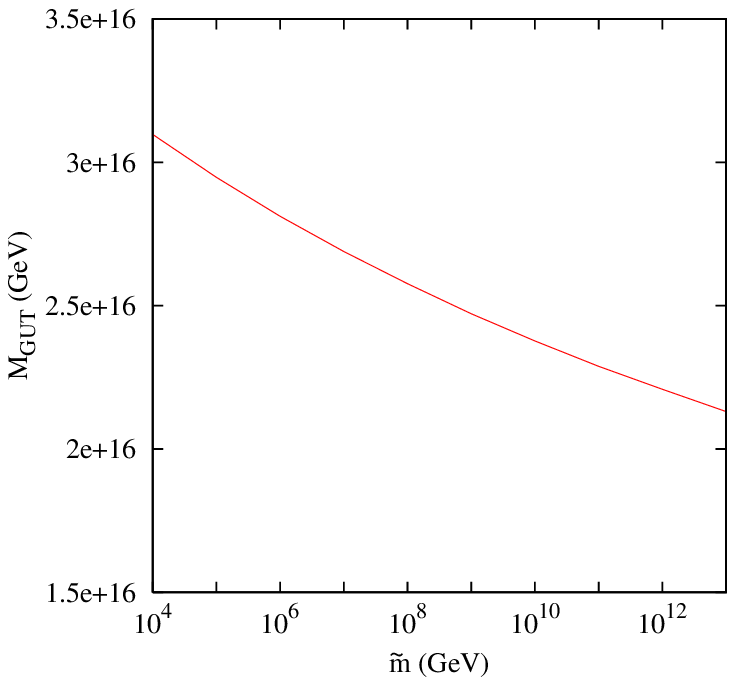}
\caption{\label{fig4}
Variations of $\alpha_{G}$ and $M_{\rm GUT}$ as functions of $\widetilde m$.
Varying $\tan\beta$ in the allowed range of Fig.\ref{fig2} does
not affect the curves.
}
\end{figure}

The fixed point behaviour of the top Yukawa coupling depends also on
gauge coupling strengths. 
The unified coupling strength $\alpha_G$ and the grand unifying scale 
$M_{GUT}$ are plotted in Fig.\ref{fig4} as functions of $\widetilde m$.
In this figure $\alpha_G$ and $M_{GUT}$ are shown to decrease with 
increasing $\widetilde m$. The effect of varying $\tan\beta$ in the allowed 
range of Fig.\ref{fig2} has been found to be negligible.
The decrease is due to the fact that the effective particle content in
split supersymmetry is smaller than in the MSSM; thus as $\widetilde m$
becomes larger, the running with split SUSY RG equations becomes longer
and the coupling constants meet at a smaller scale with a smaller unified
value. This feature has also been noticed in Ref.\cite{giudice-romanino}. 
The values of $\alpha_2$ and $\alpha_1$ at the electroweak scale are 
$\sim$ 0.0335 and 0.0168, respectively. 
An important point is that $M_{GUT}$, decreasing with 
$\widetilde m$, poses no threat to the longevity of the proton here since,
as pointed out in Ref.\cite{giudice-romanino}, dimension five and six
operators -- relevant to proton decay -- continue to remain
suppressed. We have also considered the variation of the QCD coupling
$\alpha_s(M_Z)$ with $\widetilde m$ with a result 
not very different from that of 
Ref.\cite{giudice-romanino}.    

\begin{figure}
\includegraphics{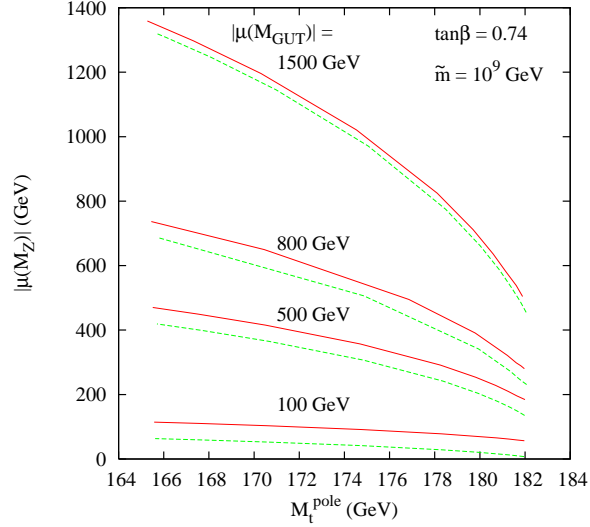}
\caption{\label{fig5}
Variation of $|\mu(M_Z)|$ near the top-quark fixed point for different values
of $|\mu(M_{GUT})|$. Here, $\widetilde m = 10^9$ GeV and $\tan\beta$=
0.74 . The common gaugino mass at the GUT scale ($M_{1/2}$) is taken to be
300 GeV. Solid/Red lines correspond to negative $\mu$ and the 
dashed/green lines correspond to positive $\mu(M_{GUT})$.}
\end{figure}

\begin{figure}
\includegraphics{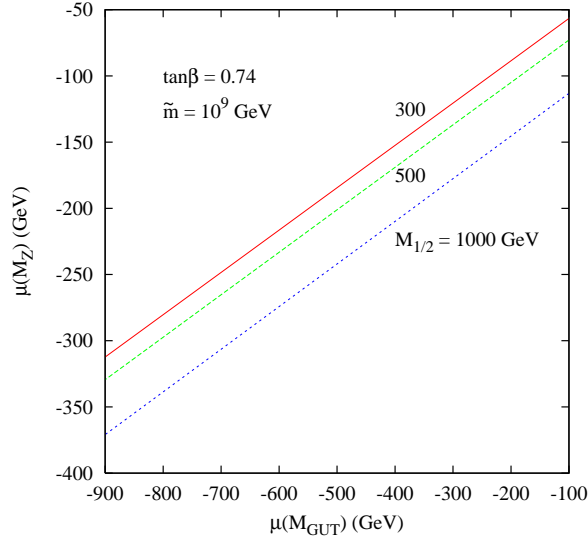}
\caption{\label{fig6}
Variation of $\mu(M_Z)$ near the top-quark fixed point for different values
of common gaugino mass as a function of $\mu(M_{GUT})$. Here, 
$\widetilde m = 10^9$ GeV and $\tan\beta$= 0.74.}
\end{figure}

Consider now how other parameters, such as $\mu(M_Z)$ and gaugino
coupling strengths vary with $M_t$ in the neighborhood of the fixed
point value.  Fig. \ref{fig5} shows precisely such a variation in
$\mu(M_Z)$, plotted vs.  $M_t^{pole}$, for various choices of 
$\mu(M_{\rm GUT})$ and $\widetilde m$= $10^9$ GeV. The common gaugino mass at
the GUT scale has been taken to be 300 GeV and $\tan\beta$ = 0.74. Running 
with RGE's brings $\mu(M_{\rm GUT})$ to $\mu(M_Z)$. Evident from the 
figure is the fact that for this choice of $\tan\beta$ = 0.74, the fixed point 
value of the top pole mass ($\sim$ 182 GeV) is within the 1$\sigma$ 
experimental band and we should look into the variation of $\mu(M_Z)$ in 
this region of the parameter space. We can see that near the Landau pole 
the change in $|\mu(M_Z)|$ is sharp for larger values$^3$ of 
$|\mu(M_{GUT})|$, less so when the latter is closer to the EW scale.
The value of $\mu(M_Z)$ can be determined (possibly along with 
$\tan\beta$) from the measurements of neutralino and
chargino masses \cite{choi} at lepton colliders. Hence, with a precise
measurement of the top mass and with the measured value of $|\mu(M_Z)|$
and $\tan\beta$, one can predict the value of $\mu(M_{\rm GUT})$ from 
the above plots for a given $\widetilde m$. Of course, it is true that this 
figure is drawn for a particular value of the common gaugino mass. In 
order to get some idea of the dependence of $\mu(M_Z)$ on the gaugino 
mass we have shown in Fig.\ref{fig6} the variation in $\mu(M_Z)$ as 
a function of $\mu(M_{GUT})$ at the fixed point for three different 
values of $M_{1/2}$ and for the same choice of $\tan\beta$ and 
$\widetilde m$ as in Fig.\ref{fig5}. We have also checked that the gaugino
mass parameters $M_2$ and $M_1$ show little variation as functions of
top mass near the fixed point which we do not show here.  

\begin{figure}
\includegraphics{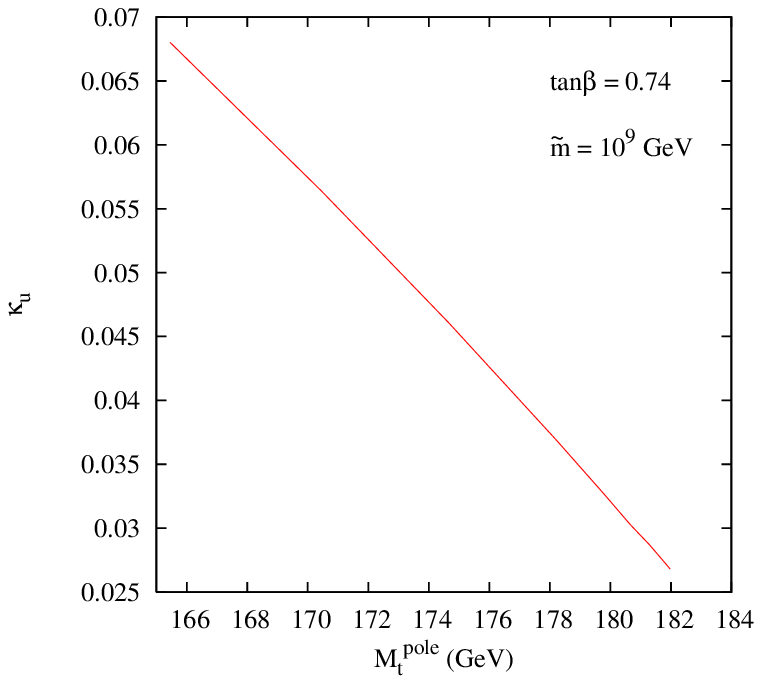}
\includegraphics{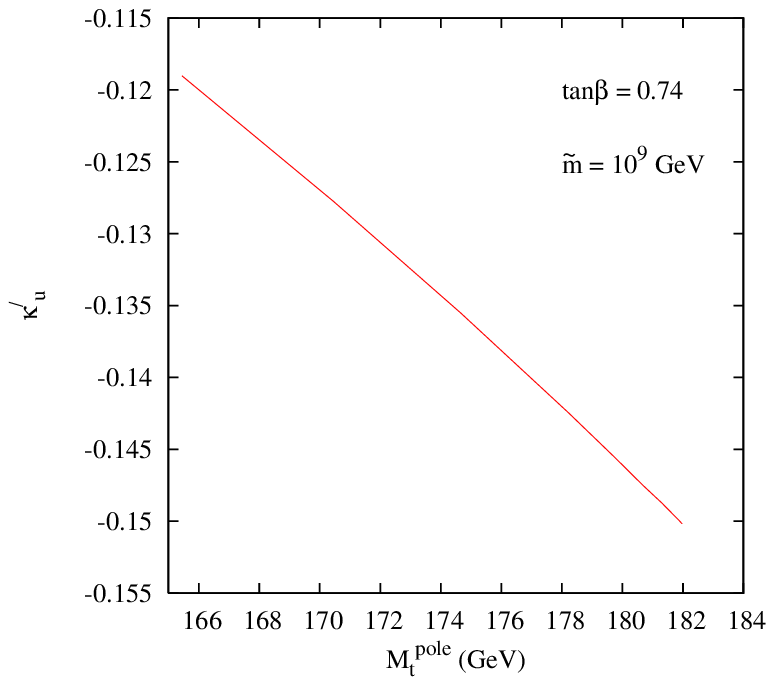}
\includegraphics{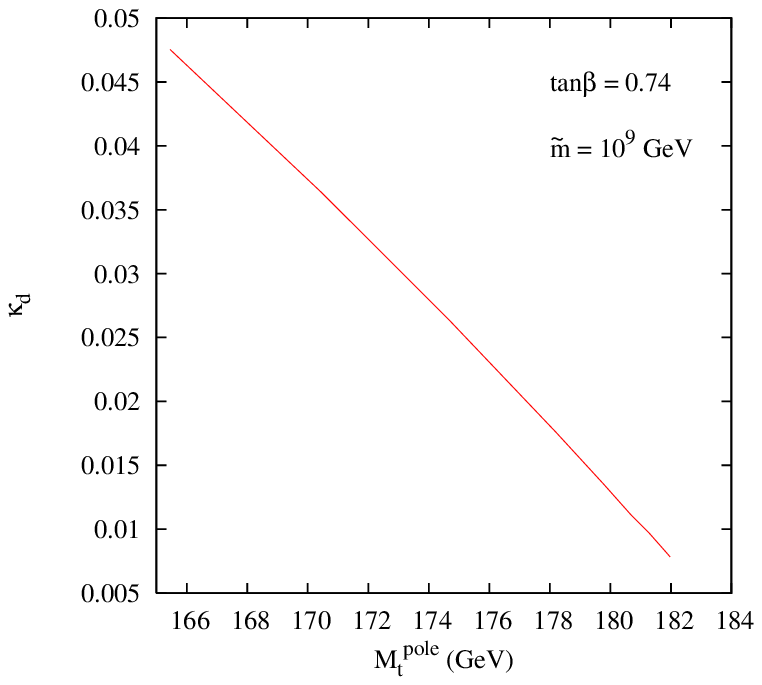}
\includegraphics{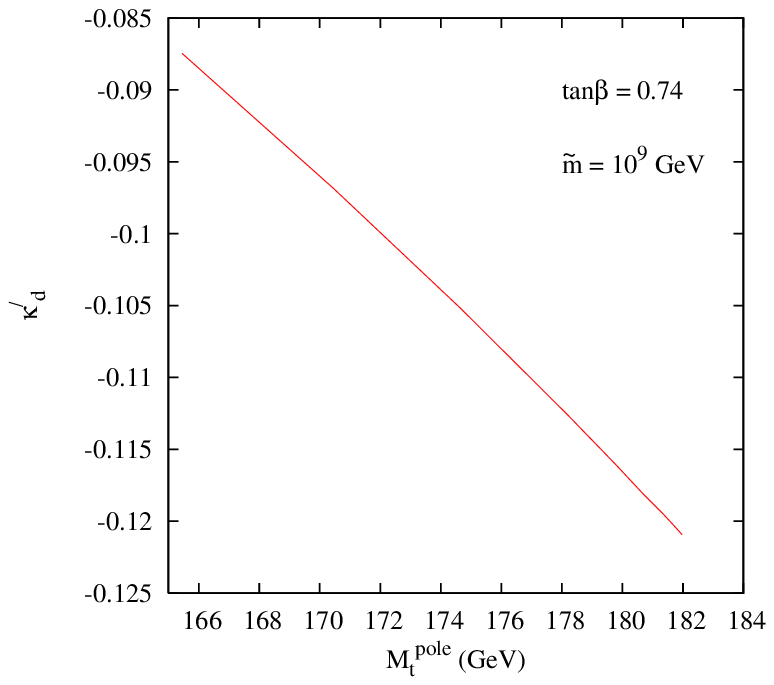}
\caption{\label{fig7}
Variation of the gaugino couplings (defined in Eqn. \ref{kappa} and
\ref{kappaprime}) near the top-quark fixed point for $\tan\beta$ = 0.74 for 
which the fixed point value of $M_t^{pole}$ = 181.9 GeV. Here, $\tilde
m$ = $10^9$ GeV.}
\end{figure}

Another important split SUSY prediction is the inequality of the gauge and 
gaugino coupling strengths below the scale $\widetilde m$. This effect is
large on account of the ultraheaviness of the sfermions and can be
detected in collider experiments involving gaugino production. 
The part of the Lagrangian, containing the gaugino couplings, 
can be written in the notation of Ref. \cite{giudice-romanino}
\bea
\label{gaugino-coupling}
{\cal L}_{gaugino-int.} = {{h^\dagger} \over \sqrt{2}} ({\tilde
g}_u \sigma^a {\tilde W}^a + {\tilde g^\prime_u}{\tilde B}){\tilde H}_u
+{{h^T \epsilon} \over \sqrt{2}}(-{\tilde g}_d \sigma^a {\tilde W}^a +
{\tilde g^\prime_d}{\tilde B}){\tilde H}_d + h.c.
\eea
\noindent Here ${\tilde H}_{u,d}$ are the `up,down type' higgsino 
fields, ${\tilde W}$ and ${\tilde B}$ are the Wino and the Bino 
respectively, $h$ is the Higgs field and $\epsilon = i \sigma_2$. 
The boundary conditions of the gaugino couplings at $\widetilde m$ are 
as follows :
\bea
{\tilde g}_u(\widetilde m) = g(\widetilde m)~{\rm sin}\beta, 
~~{\tilde g}_d(\widetilde m) = g(\widetilde m)~{\rm cos}\beta 
\eea
\bea
{\tilde g^\prime_u}(\widetilde m) = g^\prime(\widetilde m)~{\rm sin}\beta, 
~~{\tilde g^\prime_d}(\widetilde m) = g^\prime(\widetilde m)~{\rm cos}\beta. 
\eea
These couplings are then evolved to the electroweak scale using 
the renormalization group equations given in the Appendix. It is 
interesting to see the behaviour of these couplings near the infrared 
fixed point of the top mass. Following Ref.\cite{collider}, one can 
define `anomalous' gaugino couplings $\kappa_{u,d}$, $\kappa^\prime_{u,d}$ 
by the following equations, 
\bea
\kappa_u = 1 - {{\tilde g}_u \over {g~{\rm sin}\beta}},\;\;
\kappa_d = 1 - {{\tilde g}_d \over {g~{\rm cos}\beta}},
\label{kappa}
\eea 
\bea
\kappa^\prime_u = 1 - {{\tilde g}^\prime_u \over {g^\prime~{\rm sin}
\beta}},\;\;
\kappa^\prime_d = 1 - {{\tilde g}^\prime_d \over {g^\prime~{\rm cos}
\beta}}.
\label{kappaprime}
\eea 
The behaviour of these anomalous gaugino couplings near the infrared
fixed point top mass is shown in Fig.\ref{fig7}. 
Measurements of gaugino couplings $\tilde g$ and gauge couplings $g$
lead to the determination of $\widetilde m$, if $\tan\beta$ is known:
according to Eqs. (10) and (11), the couplings $\kappa_{u,d}$
and $\kappa^{'}_{u,d}$vanish at the scale $\widetilde m$.

\section{Conclusion}
In this paper we have studied the infra-red fixed point behaviour of the top
Yukawa coupling and its associated phenomenology in split supersymmetry.
In the fixed point scenario we find that only a thin band of the
$\tan\beta-\widetilde m$ plane is allowed. This is a combined effect of the 
experimental limits in the measurement of the top mass and the position 
of the Landau pole. This observation makes the infrared fixed point
scenario heavily disfavored in the context of split supersymmetry, since
it requires additional unnatural cancellation of parameters (in usual gauge, 
gravity or anomaly mediated supersymmetry breaking) in order to keep 
$\tan\beta$ within the allowed limits. One should, however, note that
such smaller values of $\tan\beta$ can possibly be obtained in the context 
of direct mediation of supersymmetry breaking with D-terms. Even if one does 
not assume the exact fixed point value for the top mass, there is still 
a lower limit on the parameter $\tan\beta$ as a function of $\widetilde m$, 
which can be less than unity for large values of $\widetilde m$. The LEP 
constraint that the Higgs must be heavier than 114.4 GeV puts additional 
restriction on minimal split SUSY parameters. We have studied various 
couplings as well as the value of the grand unifying scale in this scenario 
and, in particular, have drawn attention to the very interesting  
behaviour of the higgsino mass parameter $\mu(M_Z)$ near the the fixed
point. We have also discussed the variations in the gaugino coupling 
strengths $\tilde g_{u,d}$, $\tilde g'_{u,d}$ and of the Higgs quartic
self coupling $\lambda$, near the fixed point.  

{\bf Note added in Proof:} After this work was submitted, we saw a paper by Delgado
and Giudice (hep-ph/0506217) which claims to have excluded the top-mass fixed point
solution in split supersymmetry incorporated within an ${\rm SU}$(5) GUT by assuming
the corresponding boundary conditions for the soft scalar masses and by requiring the
absence of charge and color violating minima.

\begin{acknowledgments}
We thank JoAnne Hewett for helpful discussions on split supersymmetry. 
P.R. acknowledges the hospitality of the Helsinki Institute of
Physics. This work is supported by the Academy of Finland 
(Project numbers 104368 and 54023). 
\end{acknowledgments}

\appendix*
\section{}

In this appendix we have written down the renormalization group
equations for split supersymmetry which are taken from Ref.
\cite{giudice-romanino} but with the notations we have used in our
numerical calculations. 

\noindent {\em Evolution between $M_{\rm GUT}$ and $\widetilde m$}  \\
The 2-loop renormalization group equations for the gauge couplings are
given by 

\bea
{d{\tilde \alpha}_i \over dt} &=& -b_i {\tilde \alpha}_i^2 - {\tilde
\alpha}_i^2 [\sum_{j=1}^3 B_{ij}{\tilde \alpha}_j - (d^t_i
Y_t + d^b_i Y_b + d^\tau_i Y_\tau)],       
\eea
\noindent where $t = 2 \ln {M_{\rm GUT} \over Q}$ and $Q$ is the
renormalization scale. ${\tilde \alpha_i} = 
{\left({g_i \over {4\pi}}\right)}^2$, ${\rm Y}_{t,b,\tau} = 
{\left({\lambda_{t,b,\tau} \over {4\pi}}\right)}^2$£. We have used 
the GUT normalization condition $g^2_1 = (5/3)g^{\prime 2}$. The 
$\beta$-function coefficients are given by 
\bea
b &=& \left({33 \over 5}, 1, -3\right),\;\;
B = \left(\matrix{199 \over 24 & 27 \over 5 & 88 \over 5 \cr
9 \over 5 & 25 & 24 \cr
11 \over 5 & 9 & 14 \cr}\right) 
\eea
\bea
d^t = \left({26 \over 5}, 6, 4\right),\;\;
d^b = \left({14 \over 5}, 6, 4\right),\;\;
d^\tau = \left({18 \over 5}, 2, 0\right)
\eea
The equations for the Yukawa couplings at the one loop level are given
by
\bea
{dY_t \over dt} &=& Y_t\left({16 \over 3} {\tilde \alpha_3} + 3 {\tilde
\alpha_2} + {13 \over 15} {\tilde \alpha_1}\right)-6 Y^2_t - Y_t Y_b
\eea
\bea
{dY_b \over dt} &=& Y_b\left({16 \over 3} {\tilde \alpha_3} + 3 {\tilde
\alpha_2} + {7 \over 15} {\tilde \alpha_1}\right)-6 Y^2_b - Y_t Y_b
-Y_b Y_\tau
\eea
\bea
{dY_\tau \over dt} &=& Y_\tau\left(3 {\tilde \alpha_2} + 
{9 \over 5} {\tilde \alpha_1}\right)-4 Y^2_\tau - 3 Y_\tau Y_b
\eea
\noindent At the one loop level the equations for the gaugino masses and
$\mu$ are given by 
\bea
{dM_i \over dt} &=& -b_i {\tilde \alpha_i}M_i
\eea
\bea
{d\mu \over dt} &=& \left[{3 \over 2}{\tilde \alpha_2} + {3 \over
10}{\tilde \alpha_1} - {3 \over 2}Y_t - {3 \over 2}Y_b - 
{1 \over 2}Y_\tau\right]\mu  
\eea

\noindent {\em Evolution between $\widetilde m$ and max ($m_t$, $M_{\tilde
\chi^0_1}$)} \\
Now, 
\bea
{d{\tilde \alpha}_i \over dt} &=& -b_i {\tilde \alpha}_i^2 - {\tilde
\alpha}_i^2 [\sum_{j=1}^3 B_{ij}{\tilde \alpha}_j -  
\{d^t_i Y^\prime_t + d^b_i Y^\prime_b + d^\tau_i Y^\prime_\tau 
- d^W_i({\tilde Y_u} + {\tilde Y_d}) - d^B_i({\tilde Y^\prime_u} 
+ {\tilde Y^\prime_d}) \}], \nonumber \\      
\eea
\noindent where 
\bea
d^W = \left({9 \over 20}, {11 \over 4}, 0\right),\;\;
d^B = \left({3 \over 20}, {1 \over 4}, 0\right),\;\;
\eea
and \\
\bea
b = \left({9 \over 2}, -{7 \over 6}, -5\right),\;\;
B =\left(\matrix{104 \over 25 & 18 \over 5 & 44 \over 5 \cr
6 \over 5 & 106 \over 3 & 12 \cr
11 \over 10 & 9 \over 2 & 22 \cr}\right) 
\eea
\bea
d^t = \left({17 \over 10}, {3 \over 2}, 2\right),\;\;
d^b = \left({1 \over 2}, {3 \over 2}, 2\right),\;\;
d^\tau = \left({3 \over 2}, {1 \over 2}, 0\right),\;\;
\eea
\noindent ${\tilde {\rm Y}_u}$, ${\tilde {\rm Y}_d}$, 
${\tilde {\rm Y}^\prime_u}$, ${\tilde {\rm Y}^\prime_d}$ are 
defined generically as ${\tilde {\rm Y}} = {{\tilde g}^2 \over 
{(4\pi)}^2}$ where the gaugino couplings (${\tilde g}$'s) are 
defined in Eqn.(\ref{gaugino-coupling}) and ${\rm Y}^\prime_{t,b,\tau} 
= {\left({h_{t,b,\tau} \over {4\pi}}\right)}^2$ with $h_t$ and 
$\lambda_t$ are related by Eqn.(\ref{eqn4}) and $h_{b,\tau}(\widetilde m) 
= \lambda^*_{b,\tau}(\widetilde m)cos\beta$. 

\noindent Below the scale $\widetilde m$ the renormalization group equations 
of the Yukawa couplings at the one loop level are given by
\bea
{dY^\prime_t \over dt} &=& 3Y^\prime_t\left({8 \over 3} {\tilde \alpha_3} 
+ {3 \over 4} {\tilde \alpha_2} + {17 \over 60} {\tilde \alpha_1}\right)
- {1 \over 2} Y^\prime_t(9Y^\prime_t+3Y^\prime_b+2Y^\prime_\tau+3{\tilde
Y_u}+3{\tilde Y_d} +{\tilde Y^\prime_u}+{\tilde Y^\prime_d}) \nonumber
\\
\label{eq-a.13}
\eea
\bea
{dY^\prime_b \over dt} &=& 3Y^\prime_b\left({8 \over 3} {\tilde \alpha_3} 
+ {3 \over 4} {\tilde \alpha_2} + {1 \over 12} {\tilde \alpha_1}\right)
- {1 \over 2} Y^\prime_b(3Y^\prime_t+9Y^\prime_b+2Y^\prime_\tau+3{\tilde
Y_u}+3{\tilde Y_d} +{\tilde Y^\prime_u}+{\tilde Y^\prime_d}) \nonumber
\\
\eea
\bea
{dY^\prime_\tau \over dt} &=& 3Y^\prime_\tau\left({3 \over 4} 
{\tilde \alpha_2} + {3 \over 4} {\tilde \alpha_1}\right)
- {1 \over 2} Y^\prime_\tau(6Y^\prime_t+6Y^\prime_b+5Y^\prime_\tau+3{\tilde
Y_u}+3{\tilde Y_d} +{\tilde Y^\prime_u}+{\tilde Y^\prime_d})
\eea
The gaugino mass equations are (including next-to-leading order
corrections)
\bea
{dM_3 \over dt} &=&  9{\tilde \alpha_3}M_3 (1+c_{\tilde g}{\tilde
\alpha_3}),
\eea
where $c_{\tilde g}$ = 38/3 in ${\overline {MS}}$ and $c_{\tilde g}$ 
= 10 in ${\overline {DR}}$.
\bea
{dM_2 \over dt} &=&  6({\tilde \alpha_2} - {1 \over 2} {\tilde Y_u} - {1
\over 2} {\tilde Y_d})M_2 -2\sqrt{{\tilde Y_u} {\tilde Y_d}} \mu
\eea
\bea
{dM_1 \over dt} &=&  - {1 \over 2} ({\tilde Y^\prime_u} + 
{\tilde Y^\prime_d})M_1 -2\sqrt{{\tilde Y^\prime_u} {\tilde Y^\prime_d}} 
\mu
\eea
The renormalization group equation for the $\mu$ parameter below the
scale $\widetilde m$ is given by
\bea
{d\mu \over dt} &=& \left[{9 \over 4}\left({{\tilde \alpha_1} \over 5} +
{\tilde \alpha_2}\right) - {3 \over 8} ({\tilde Y_u} + {\tilde Y_d}) 
- {1 \over 8} ({\tilde Y^\prime_u} + {\tilde Y^\prime_d})\right]\mu
-{3 \over 2} \sqrt{{\tilde Y_u}{\tilde Y_d}}M_2 - {1 \over 2}
\sqrt{{\tilde Y^\prime_u}{\tilde Y^\prime_d}}M_1 \nonumber \\
\eea
The equations for the gaugino couplings are given by 
\bea
{d{\tilde Y_u} \over dt} &=& 3 {\tilde Y_u} \left({11 \over 4} 
{\tilde \alpha_2} + {3 \over 20} {\tilde \alpha_1}\right) - {1 \over 4}
{\tilde Y_u}(5{\tilde Y_u}-2{\tilde Y_d}+{\tilde Y^\prime_u})
-{({\tilde Y_u}{\tilde Y_d}{\tilde Y^\prime_u}{\tilde
Y^\prime_d})}^{1/2}
\nonumber
\\[1.2ex]
&&-{1 \over 2}{\tilde Y_u}(6Y^\prime_t+6Y^\prime_b+2Y^\prime_\tau+
3{\tilde Y_u}+3{\tilde Y_d}+{\tilde Y^\prime_u}+{\tilde Y^\prime_d})
\eea
\bea
{d{\tilde Y^\prime_u} \over dt} &=& 3{\tilde Y^\prime_u} \left({3 \over 4} 
{\tilde \alpha_2} + {3 \over 20} {\tilde \alpha_1}\right) - {3 \over 4}
{\tilde Y^\prime_u}({\tilde Y^\prime_u}+2{\tilde Y^\prime_d}+
{\tilde Y_u})-3{({\tilde Y_u}{\tilde Y_d}{\tilde Y^\prime_u}{\tilde
Y^\prime_d})}^{1/2}
\nonumber
\\[1.2ex]
&&-{1 \over 2}{\tilde Y^\prime_u}(6Y^\prime_t+6Y^\prime_b+2Y^\prime_\tau+
3{\tilde Y_u}+3{\tilde Y_d}+{\tilde Y^\prime_u}+{\tilde Y^\prime_d})
\eea
\bea
{d{\tilde Y_d} \over dt} &=& 3 {\tilde Y_d} \left({11 \over 4} 
{\tilde \alpha_2} + {3 \over 20} {\tilde \alpha_1}\right) - {1 \over 4}
{\tilde Y_d}(-2{\tilde Y_u}+5{\tilde Y_d}+{\tilde Y^\prime_d})
-{({\tilde Y_u}{\tilde Y_d}{\tilde Y^\prime_u}{\tilde
Y^\prime_d})}^{1/2}
\nonumber
\\[1.2ex]
&&-{1 \over 2}{\tilde Y_d}(6Y^\prime_t+6Y^\prime_b+2Y^\prime_\tau+
3{\tilde Y_u}+3{\tilde Y_d}+{\tilde Y^\prime_u}+{\tilde Y^\prime_d})
\eea
\bea
{d{\tilde Y^\prime_d} \over dt} &=& 3{\tilde Y^\prime_d} \left({3 \over 4} 
{\tilde \alpha_2} + {3 \over 20} {\tilde \alpha_1}\right) - {3 \over 4}
{\tilde Y^\prime_d}({\tilde Y^\prime_d}+2{\tilde Y^\prime_u}+
{\tilde Y_d})-3{({\tilde Y_u}{\tilde Y_d}{\tilde Y^\prime_u}{\tilde
Y^\prime_d})}^{1/2}
\nonumber
\\[1.2ex]
&&-{1 \over 2}{\tilde Y^\prime_d}(6Y^\prime_t+6Y^\prime_b+2Y^\prime_\tau+
3{\tilde Y_u}+3{\tilde Y_d}+{\tilde Y^\prime_u}+{\tilde Y^\prime_d})
\eea

Now, the evolution equation for the Higgs quartic coupling $\lambda$ is
\bea
{d{\tilde \lambda} \over dt} &=& -6{\tilde \lambda}^2 - {1 \over 2}{\tilde
\lambda} [-9\left({1 \over 5}{\tilde \alpha_1}+{\tilde \alpha_2}\right)
+6({\tilde Y_u}+{\tilde Y_d})+2({\tilde Y^\prime_u}+{\tilde Y^\prime_d})
+12 Y^\prime_t + 12 Y^\prime_b + 4Y^\prime_\tau ]
\nonumber
\\[1.2ex]
&&-{9 \over 4} \left({1 \over 2} {\tilde \alpha_2}^2 + {3 \over 50} 
{\tilde \alpha_1}^2 + {1 \over 5}{\tilde \alpha_1} {\tilde \alpha_2}\right)
+{5 \over 2}({\tilde Y_u}^2+{\tilde Y_d}^2)+{\tilde Y_u}{\tilde Y_d}
+{1 \over 2}{({\tilde Y^\prime_u}+{\tilde Y^\prime_d})}^2
\nonumber
\\[1.2ex]
&&+ {(\sqrt{{\tilde Y_u}{\tilde Y^\prime_u}}+\sqrt{{\tilde Y_d}{\tilde
Y^\prime_d}})}^2+6Y^{\prime 2}_t+6Y^{\prime 2}_b +2Y^{\prime 2}_\tau, 
\label{eq-a.24}
\eea
\noindent where $\tilde \lambda = {\lambda \over {(4\pi)}^2}$.

{\underline {Caution:}} If $M_{\tilde \chi^0_1} > m_t$, in the evolution
from $M_{\tilde \chi^0_1}$ to $m_t$ of the gauge couplings  
\bea
b = \left({41 \over 10}, -{19 \over 6}, -7\right),\;\;
B =\left(\matrix{109 \over 50 & 27 \over 10 & 44 \over 5 \cr
9 \over 10 & 35 \over 6 & 12 \cr
11 \over 10 & 9 \over 2 & -26 \cr}\right) 
\eea
\bea
d^t = \left({17 \over 10}, {3 \over 2}, 2\right),\;\;
d^b = \left({1 \over 2}, {3 \over 2}, 2\right),\;\;
d^\tau = \left({3 \over 2}, {1 \over 2}, 0\right),\;\;
d^W=0=d^B.
\eea


\begin{thebibliography}{99}
\bibitem{arkani-dimo}
N. Arkani-Hamed and S. Dimopoulos, hep-th/0405159.

\bibitem{giudice-romanino}
G.F. Giudice and A. Romanino, Nucl. Phys. {\bf B699}, 65(2004)
(hep-ph/0406088).

\bibitem{dark-matter}
A. Pierce, Phys. Rev. D {\bf 70}, 075006(2004) (hep-ph/0406144).

\bibitem{one-loop}
A. Arvanitaki, C. Davis, P. W. Graham and J. G. Wacker, hep-ph/0406034.

\bibitem{aspects}
N. Arkani-Hamed, S. Dimopoulos, G.F. Giudice and A. Romanino,
Nucl. Phys. {\bf B709}, 3 (2005) (hep-ph/0409232).

\bibitem{spectrum}
B. Mukhopadhyaya and S. SenGupta, Phys. Rev. D {\bf 71}, 035004 (2005)
(hep-th/0407225).

\bibitem{collider}
W. Kilian, T. Plehn, P. Richardson and E.  Schmidt, Eur. Phys. J. 
{\bf C39}, 229 (2005) (hep-ph/0408088).

\bibitem{zhu}
S-H. Zhu, Phys. Lett. {\bf B604}, 207 (2004) (hep-ph/0407072).

\bibitem{2-loop}
M. Binger, hep-ph/0408240.

\bibitem{axion}
V. Barger, C-W. Chiang, J. Jiang and T. Li, Nucl. Phys. {\bf B705}, 
71 (2005) (hep-ph/0410252).

\bibitem{variation}
R. Mahbubani, hep-ph/0408096.

\bibitem{r-parity}
S. K. Gupta, P. Konar and B. Mukhopadhyaya, Phys. Lett. {\bf B606}, 
384 (2005) (hep-ph/0408296).

\bibitem{neutrino}
E. J. Chun and S. C. Park, J. High Energy Phys. {\bf
01}, 009 (2005) (hep-ph/0410242).

\bibitem{neutralino-dark}
A. Masiero, S. Profumo and P. Ullio, Nucl. Phys. {\bf B712}, 86 (2005) 
(hep-ph/0412058).

\bibitem{diaz-perez}
M.A. D\'iaz and P. F. Perez, J. Phys. {\bf G31}, 1 (2005) 
(hep-ph/0412066).

\bibitem{hewett}
J. L. Hewett, B. Lillie, M. Masip, T.G. Rizzo, J. High Energy Phys. {\bf
09}, 070 (2004).

\bibitem{cosmic}
L. Anchordoqui, H. Goldberg, C. Nu{$\tilde n$}ez, Phys. Rev. D {\bf 71},
065014 (2005) (hep-ph/0408284).  

\bibitem{keung}
K. Cheung, W-Y. Keung, Phys. Rev. D {\bf 71}, 015015 (2005) 
(hep-ph/0408335).

\bibitem{sarkar}
U. Sarkar, hep-ph/0410104.

\bibitem{allahverdi}
R. Allahverdi, A. Jokinen, A. Mazumdar, Phys. Rev. D {\bf 71}, 043505 
(2005) (hep-ph/0410169).

\bibitem{bajc}
B. Bajc, G. Senjanovic, Phys. Lett. {\bf B610}, 80 (2005) 
(hep-ph/0411193).

\bibitem{graham}
A. Arvanitaki, P.W. Graham, hep-ph/0411376.

\bibitem{senatore}
L. Senatore, hep-ph/0412103.

\bibitem{vacuum}
A. Datta, X. Zhang, hep-ph/0412255.

\bibitem{schuster}
P.C. Schuster, hep-ph/0412263.

\bibitem{gao}
G. Gao, R.J. Oakes, J.M. Yang, hep-ph/0412356.

\bibitem{martin}
S.P. Martin, K. Tobe, J.D. Wells, hep-ph/0412356.

\bibitem{chen}
C.H. Chen, C.Q. Geng, hep-ph/0501001.

\bibitem{babu}
K.S. Babu, Ts. Enkhbat, B. Mukhopadhyaya, hep-ph/0501079.

\bibitem{drees}
M. Drees, hep-ph/0501106.

\bibitem{shinta}
S. Kasuya, F. Takahashi, hep-ph/0501240.

\bibitem{cheung}
K. Cheung, C-W. Chiang, hep-ph/0501265. 

\bibitem{hill}
B. Pendleton and G. G. Ross, Phys. Lett. {\bf B98}, 291 (1981); C. T.
Hill, Phys. Rev. D{\bf 24}, 691 (1981).

\bibitem{ibanez}
L. Iba{$\tilde {\rm n}$}ez and C. Lopez, Nucl. Phys. {\bf B233}, 
511 (1984);\\
L. Iba{$\tilde {\rm n}$}ez, C. Lopez and C. Mu{$\tilde {\rm n}$}oz, 
Nucl. Phys. {\bf B256}, 218 (1985). 

\bibitem{savoy}
A. Bouquet, J. Kaplan and C.A. Savoy, Nucl. Phys. {\bf B262}, 299 (1985).

\bibitem{carena}
M. Carena,  M. Olechowski, S. Pokorski and C.E.M. Wagner, Nucl. Phys. 
{\bf B419}, 213 (1994).

\bibitem{nath}
P. Nath, J. Wu and R. Arnowitt, Phys. Rev. D {\bf 52}, 4169(1995).

\bibitem{kobayashi}
T. Kobayashi and Y. Yamagishi, Phys. Lett. {\bf B381}, 169 (1996).

\bibitem{biswajoy}
B. Brahmachari, Mod. Phys. Lett. {\bf A12}, 1969 (1997).

\bibitem{non-univ}
D.A. Demir, hep-ph/0410056;\\
B. K\"ors and P. Nath, Nucl. Phys. {\bf B711}, 112 (2005) (hep-th/0411201).

\bibitem{polemass}
H.E. Haber, R. Hempfling and A.H. Hoang, Z. Phys. {\bf C 75}, 539(1997);
\\
H. Arason, D.J. Castano, B. Kesthelyi, S. Mikaelian, E.J. Pirad, P.
Ramond and B.D. Wright, Phys. Rev. D {\bf 46}, 3945(1992).

\bibitem{PDG}
S. Eidelman et al., Phys. Lett. {\bf B592}, 1 (2004).

\bibitem{lephiggs}
R. Barate {\it et al.}, Phys. Lett. {\bf B565}, 61 (2003).

\bibitem{choi}
S.Y. Choi, A. Djouadi, M. Guchait, J. Kalinowski, H.S. Song, P.M. Zerwas,
Eur. Phys. J. {\bf C14}, 535 (2000);
S.Y. Choi, J. Kalinowski, G. Moortgat-Pick, P.M. Zerwas,  Eur. Phys. J. 
{\bf C22}, 563 (2001); hep-ph/0202039.

\end{thebibliography}
\end{document}